\documentclass[aps,prb,floatfix,twocolumn,showpacs,superscriptaddress,reprint]{revtex4-1}
\usepackage{epsfig}
\usepackage{amsmath}
\usepackage{amsfonts}
\usepackage{amssymb}
\usepackage{mathptmx}
\usepackage{eucal}
\usepackage{bm}

\DeclareMathOperator{\Tr}{\mathop{\mathrm{Tr}}}

\DeclareMathOperator{\re}{\mathop{\mathrm{Re}}}
\DeclareMathOperator{\im}{\mathop{\mathrm{Im}}}
\newcommand{\oM}{ \overline{M} }
\newcommand{\oh}{ \overline{h} }
\newcommand{\oN}{ \overline{N} }

\newcommand{\tv}{ \widetilde{v} }
\newcommand{\tu}{ \widetilde{u} }

\newcommand{\tD}{ \widetilde{\Delta} }
\newcommand{\cG}{ \check{G} }
\newcommand{\cA}{ \check{A} }
\newcommand{\cK}{ \check{K} }
\newcommand{\cP}{ \check{P} }
\newcommand{\cQ}{ \check{Q} }
\newcommand{\cq}{ \check{q} }
\newcommand{\Eq}[1]{Eq.\! (\ref{#1})}
\newcommand{\Eqs}[1]{Eqs.\! (\ref{#1})}

\begin{document}

\title{Current-voltage characteristics of asymmetric
double-barrier Josephson junctions}

\author{E.~V.~Bezuglyi}
\email{eugene.bezuglyi@gmail.com}
\author{E.~N.~Bratus'}
\affiliation{B.Verkin Institute for Low Temperature Physics and Engineering,
61103 Kharkov, National Academy of Sciences, Ukraine}
\author{V.~S.~Shumeiko}
\affiliation{Chalmers University of Technology, S-41296 G\"oteborg, Sweden}

\pacs{74.50.+r, 74.45.+c}

\begin{abstract}
We develop a theory for the current-voltage characteristics of diffusive
superconductor-normal metal-super\-conductor Josephson junctions with resistive
interfaces and the distance between the electrodes smaller than the
superconducting coherence length. The theory allows for a quantitative
analytical and numerical analysis in the whole range of the interface
transparencies and asymmetry. We focus on the regime of large interface
resistance compared to the resistance of the normal region, when the
electron-hole dephasing in the normal region is significant and the finite
length of the junction plays a role. In the limit of strong asymmetry we find
pronounced current structures at the combination subharmonics of
$\Delta+\Delta_g$, where $\Delta_g$ is the proximity minigap in the normal
region, in addition to the subharmonics of the energy gap $2\Delta$ in the
electrodes. In the limit of rather transparent interfaces, our theory recovers
a known formula for the current in a short mesoscopic connector -- a
convolution of the current through a single-channel point contact with the
transparency distribution for an asymmetric double-barrier potential.
\end{abstract}

\maketitle

\section{Introduction}
\vskip -3mm

During last few years a large number of experimental researches has been done
on the proximity effect in semiconductor nanowires connected to superconducting
electrodes.\cite{
Kewo,Tinkham,Jespersen,Doh,Schoenenberger,Takahiro,Xu,simon,Abay,
KretininCM,AbayCM} Hybrid devices of the nanowires have demonstrated Andreev
subgap conductance, Josephson field effect, and Cooper-pair beam splitting.
More recently, the nanowire hybrid devices attracted new attention following
theoretical predictions of Majorana bound states in nanowire proximity
structures.

From a theory viewpoint, the majority of investigated devices fall into the
category of mesoscopic diffusive supercon\-ductor-normal metal-superconductor
(SNS) junctions with the length smaller or comparable to the superconducting
coherence length. These devices typically have about 100 conducting channels,
impurity mean free path $ \sim 30-50$ nm, and the length varying from tens to
hundreds nanometers, i.e. the Thouless energy $E_{\textrm{Th}}$ in the range of
$10-0.1$ meV. This is larger or comparable to the energy gap $\Delta$ in
superconducting Al, which is used as the electrode material. The most
interesting regime of a strong proximity effect, manifested by considerable
Josephson current, is achieved in junctions with rather transparent
nanowire-su\-per\-conductor interfaces, whose transparencies typically exceed
0.1.

The physics of the equilibrium proximity effect in such junctions is
qualitatively well understood, and a quantitative theory for the dc Josephson
transport has been developed by many authors on the basis of the quasiclassical
Greens function technique, see, e.g., Ref.~\onlinecite{GolubovReview} and
references therein.

Quantitative description of the ac Josephson effect is more challenging. The
difficulty here arises from the presence of the time dependence of the dynamics
in the normal region, in addition to the spatial inhomogeneity and
nonlinearity. The problem was found solvable in diffusive point
contacts,\cite{Averin97,AZ} where approximation of the zero contact length is
appropriate. In that case, the problem can be reduced to the single channel
coherent multiple Andreev reflection (MAR)
problem\cite{Bratus95,Averin95,Cuevas1996}. Such an approximation is not
suitable for the interpretation of the nanowire experiments, which show
pronounced length dependence of the transport characteristics.

A step towards solving this problem was made in Ref.\onlinecite{Our2011} where
the coherent MAR problem has been analytically solved for a finite-length SNS
junction with highly resistive interfaces (I), SINIS. In this model, the length
of the junction is assumed to be small, but it  cannot be put equal to zero
because of significant dwell time of quasiparticles in the normal region
confined by the strong interface barriers. As it was shown, the parameter that
defines the short junction limit in this case is $\gamma=
(R/R_N)(\Delta/E_{\textrm{Th}}) \ll 1$, rather than $\Delta/E_{\textrm{Th}} \ll
1$, where $R$ is the interface resistance and $R_N$ is the resistance of the
normal region. Therefore even if the latter condition is fulfilled, the
parameter $\gamma$ can be large, $\gamma\gg 1$. This is the most interesting
regime, the physics of which is characterized, qualitatively similar to the
long junction case, by large electron-hole dephasing in the normal region,
leading to significant length dependence of the transport.

Solution of this problem is also important for understanding the properties of
the coherent current transport in planar Nb/Al$_2$O$_3$/Al/Al$_2$O$_3$/Nb
tunnel junctions which can be used as basic elements of practical
superconducting electronics: rapid single flux quantum devices,\cite{Likh}
voltage standards,\cite{Ham} high-frequency mixers,\cite{Dier}
SQUIDs\cite{Bart} (see also a review in Ref.~\onlinecite{Kupr1999}). As a rule,
in such devices, the thickness of the Al layer is about 10 nm, and the
parameter $\gamma$ may achieve the values of the order of $10^2-10^4$.

In this paper we discuss the extension of theory developed in
Ref.\onlinecite{Our2011} to a practically important case of asymmetric
junctions, namely junctions with different interface resistances. As we show,
the asymmetry leads to a qualitative change of the IVC characteristics. In
these junctions a novel set of current features appears at subharmonics of
$\Delta+\Delta_g$, where $ \Delta_g$ is the proximity induced minigap in the
normal region.

The structure of the paper is as follows. A formal solution for the
Keldysh-Green's function equation is presented in Section II. Section III is
devoted to demonstration of computation of equilibrium Josephson current using
the found solution. In Section IV, a general non-equilibrium case is
considered, and the dc current-voltage characteristics are computed in Section
V; there we present the numerical results and analytical expressions for the
partial MAR currents and the excess current.

\section{Construction of approximate solution }
\vskip -3mm

We describe our junction with the diffusive equation\cite{LO} for the
Keldysh-Green's function $\check{G}(x,t_1,t_2)$ in the normal region ($-d < x <
d$), assuming $\hbar = 1$,
\begin{equation}\label{EqKeldysh}
\left[\sigma_z \hat{E}, \cG\right] = i\mathcal{D}\partial_x \left(\cG
\partial_x \cG\right), \quad \cG^2 = 1,\quad \cG =
\begin{pmatrix} \hat{g}^R & \hat{G}^K \\ 0 & \hat{g}^A
\end{pmatrix},
\end{equation}
and the boundary conditions at the normal metal-super\-con\-duc\-tor interfaces
with the resistances $R_1$ (right) and $R_2$ (left),\cite{KL}
\begin{align}\label{KL}
g_N\left(\check{G} \partial_x \check{G}\right)_{\pm d} &= \pm
(2R_{1,2})^{-1}\left[\check{G}_{\pm d}, \check{G}_{1,2}\right].
\end{align}
Here $\hat{g}^{R,A}$ are the retarded/advanced Green's functions, $\hat G^K=
\hat{g}^R \hat{f} - \hat{f} \hat{g}^A$ is the Keldysh function with the matrix
distribution function $\hat{f}$, $\mathcal{D}$ is a diffusion constant, the
kernel of the energy operator $\hat{E}$ is ${E}(t_1,t_2) = i
\partial_{t_1} \delta(t_1 -t_2)$, and `check' and `hat' denotes $4\times 4$
Keldysh and $2\times 2$ Nambu matrices, respectively. All products in
\Eq{EqKeldysh} are time convolutions, $(A B)(t_1,t_2) = \int dt
A(t_1,t)B(t,t_2) $.

The equilibrium Keldysh-Green's functions $\check{G}_{1,2}$ in the right and
left reservoirs are constructed from the local-equi\-lib\-rium Green's and
distribution functions. In $(E,t)$-rep\-re\-sen\-tation, $A(E,t)=\int {d\tau}
e^{iE\tau}A(t+\tau/2,t-\tau/2)$, they read
%
\begin{align}
&\hat{g}_{1,2}=\sigma_z u(E\pm \sigma_z eV/2)+i\exp({\pm i\sigma_z
eVt})\sigma_y v(E),\label{g}
\\
&\hat{f}_{1,2}=\tanh[(E\pm \sigma_z eV/2)/{2T}],
\\
\label{gRL} &u(E) = \frac{E}{\xi},\quad v(E) = \frac{\Delta}{\xi}, \quad
\xi^{R,A} =\sqrt{(E\pm i0)^2-\Delta^2}.
\end{align}
In \Eq{g}, we use the antisymmetric gauge of the superconducting phase, $\phi_1
= -\phi_2 = eVt$, satisfying the Josephson relation $\phi=\phi_1-\phi_2 =
2eVt$.

The electric current $I(t)$ is defined as
\begin{equation} \label{I1}
I(t) = (\pi{g_N}/{4e}) \Tr \tau_K \left(\cG \partial_x \cG\right)(t,t) ,\quad
\tau_K=\sigma_z \tau_x,
\end{equation}
where $g_N$ is the conductance of the normal region per unit length, and the
$\sigma$ and $\tau$ Pauli matrices operate in the Nambu and the Keldysh space,
respectively.

We construct an approximate solution to \Eqs{EqKeldysh} and (\ref{KL}) by
performing integration of the diffusive equation along the coordinate $x$ of
the normal region, replacing $\cG$ in the left-hand side with its spatially
averaged value $\overline \cG$ and using the boundary condition \Eq{KL},
\begin{align}\label{KeldyshAverage}
2d[\sigma_z \hat{E}, \overline \cG] =
\frac{i\mathcal{D}}{2g_N}\Bigl(\Bigl[\cG_d, \frac{\cG_1}{R_1}\Bigr] +
\Bigl[\cG_{-d}, \frac{\cG_2}{R_2}\Bigr]\Bigr).
\end{align}
In short junctions with opaque barriers, the resistance of which exceeds the
normal resistance $R_N =2d/g_N$ of the normal region, the function $\cG$ slowly
varies along the normal region,\cite{Volkov,3T} so that $\cG\approx
\overline\cG \approx \cG_d\approx \cG_{-d}$.
This approximation leads to a simplified equation for the single quantity
$\cG$,
\begin{align}\label{Eqaverage}
2d[\sigma_z \hat{E}, \cG] = \frac{i\mathcal{D}}{2g_N}\left[\cG,
\frac{\cG_1}{R_1}+\frac{\cG_2}{R_2} \right].
\end{align}
In a similar way one can get a simplified  equation for the current, taking
symmetrized value of the current at the ends of the normal region and using the
boundary condition \Eq{KL},
\begin{eqnarray}\label{Iaverage}
I(t) = \frac{\pi}{8e} \Tr \tau_K \left[\cG\,,
 \frac{\cG_1}{R_1}-\frac{\cG_2}{R_2} \right ](t,t).
\end{eqnarray}

The simplified Green's function equation (\ref{Eqaverage}) and equation for the
current (\ref{Iaverage}) can be written in a more compact form by introducing
quantities
\begin{align}\label{Gpm}
&\cA = \cG_+ -i\sigma_z \tau \hat{E},\quad \cG_\pm = \frac{1}{2} (r_1\cG_1 \pm
r_2 \cG_2),
\\
\label{defA} &r_{1,2}= \frac{R}{R_{1,2}}, \;\; \frac{1}{R} =
\frac{1}{2}\left(\frac{1}{R_1} + \frac{1}{R_2}\right), \;\;\gamma = \tau\Delta
= \frac{R}{R_N} \frac{\Delta}{E_{\textrm{Th}}}.
\end{align}
The parameter $\gamma$ introduced in \Eq{defA} quantifies the effect of the
electron-hole dephasing, and $\tau = E_{\textrm{Th}}^{-1}R/R_N$, where the
Thouless energy is defined as $E_{\textrm{Th}}=\mathcal{D}/(2d)^2$,
characterizes the dwell time. In these notations, we obtain the equations
\begin{align}\label{CommEq}
&[\cA\,,\cG] = 0,
\\
&I(t) = \frac{\pi}{8eR} \Tr \tau_K [\cG,\cG_-](t,t). \label{Current}
\end{align}

Following Refs.~\onlinecite{Samuel} and \onlinecite{Solution}, we write a
formal solution to \Eq{CommEq} which obeys the commutation relation in
\Eq{CommEq} and the normalization condition in \Eq{EqKeldysh},
\begin{equation} \label{SolSqrt}
\cG = \cA\bigl/\sqrt{\cA^2}.
\end{equation}
A constructive form of \Eq{SolSqrt} appropriate for the analysis of a
nonstationary regime can be obtained by means of the integral
representation\cite{Our2011}
\begin{equation} \label{FormalSol}
\cG  = \frac{1}{\pi}\int_{-\infty}^\infty
d\lambda\, \cK(\lambda), \qquad \cK(\lambda)=(\cA+i\lambda)^{-1},
\end{equation}
where the integral is assumed to be taken in symmetric limits which
simultaneously turn to $\pm\infty$. Then \Eq{Current} reads
\begin{equation} \label{Current2}
I(t) =  \int_{-\infty}^\infty  \frac{d\lambda}{8 eR} \Tr \tau_K
\left[\cK(\lambda), \cG_- \right](t,t).
\end{equation}

Equations (\ref{FormalSol}) and (\ref{Current2}) are the main technical result
of the paper; they describe short asymmetric double-barrier SNS junctions for
all values of parameter $\gamma$. In what follows we will apply these equations
for calculation of the dc current-voltage characteristics.

The chosen form for the solution is justified by the limit of vanishing
dephasing parameter, $\gamma =0$, when \Eq{Current2} reduces to a known
universal formula for the current through a short
connector.\cite{Nazarov99,BelzigNazarov} Indeed, in this case, reducing the
integral in \Eq{Current2} to the positive axis, we have $\cK(\lambda)=2\cG_+
(\cG^2_+ + \lambda^2)^{-1}$, and then, after simple algebra, we obtain the
commutator in \Eq{Current2} in terms of the functions $\cG_{1,2}$:
\begin{align} \label{comm}
&\left[\check{K}(\lambda),\check{ G}_{-}\right] = \frac{1}{2}\frac{r_1 r_2
\left[\cG_2 ,\cG_1\right]}{\lambda^2 +\frac{1}{4}\left(r_1^2+r_2^2+r_1 r_2
\left\{\cG_1,\cG_2 \right \}\right)}.
\end{align}
Substituting \Eq{comm} to \Eq{Current2}, using the equality $r_1+r_2=2$, and
introducing the transparency variable $D=r_1 r_2/(\lambda^2 +1)$,  we arrive at
a convolution of a non-resonant single-channel current with the transparency
distribution $\rho(D)$ for an asymmetric double-barrier
junction,\cite{Naz2B,Naz2000}
\begin{align} \label{Nazarov}
&I(t) = \frac{\pi}{4 eR_T}\int_0^{D_{max}}  \Tr \tau_K \frac{D\rho(D)
\bigl[\check{G}_2,\check{G}_1 \bigr] dD} {1 +
\frac{D}{4}\bigl(\bigl\{\check{G}_1,\check{G}_2\bigr\}-2\bigr)}(t,t),
\\
&\rho(D) = \frac{1}{\pi D^{3/2}\sqrt{D_{max}-D}}, \quad  D_{max}= r_1 r_2 =
\frac{4R_1 R_2}{R_T^2}. \label{rhoD}
\end{align}
where $R_T = R_1+R_2$ is the net resistance of the tunnel barriers.

\section{Equilibrium Josephson current and the minigap function}
\vskip -3mm

Prior to the discussion of a general nonequilibrium case, it is instructive to
demonstrate how to use \Eqs{FormalSol} and \eqref{Current2} for evaluation of
the equilibrium Josephson current. In this case, the distribution function is
equilibrium, $f = f_2 = f_1 = \tanh(E/2T)$, and we need to calculate only the
Green's functions. In the reservoirs, they are given by $\hat{g}_{1,2} =
\sigma_z u + i \exp(\pm i\sigma_z \phi/2) \sigma_y v$; the solution for the
Green's function $\hat{g}$ in the normal region has the form of \Eq{FormalSol}
with the diagonal (retarded and advanced) component $\hat{A}_g$ of the full
matrix $\cA$:
\begin{align} \label{A}
&\hat{A}_g = \sigma_z (u - i\tau E) + iv[\sigma_y\cos(\phi/2) +\sigma_x \kappa
\sin(\phi/2)],
\\
&\hat{A}_g^2 = (u - i\tau E)^2 - v^2 \eta^2 \nonumber
,\quad \eta^2(\phi) = \cos^2 \frac{\phi}{2} +\kappa^2
\sin^2\frac{\phi}{2},\nonumber
\end{align}
where $\kappa= (R_2-R_1)/( R_2+R_1)$. As the result, we obtain
\begin{align}
&\hat{g} = {\hat{A}_g}\Bigl/{\sqrt{\hat{A}_g^2}}= \sigma_z \tu + i\tv
\exp(i\sigma_z\Phi) \sigma_y, \label{gequilib}
\\
&\tu = \frac{E}{\sqrt{E^2 - \tD^2(E,\phi)}},\quad \tv =
\frac{\tD(E,\phi)}{\sqrt{E^2 - \tD^2(E,\phi)}},\label{uv}
\\
&\tD(E,\phi) = \frac{\Delta \eta(\phi)}{1-i\gamma/v(E)}, \quad \Phi(\phi) =
\arctan\Bigl(\kappa \tan\frac{\phi}{2}\Bigr).\label{tD}
\end{align}

According to \Eq{uv}, the minigap $\Delta_g(\phi)$ in the spectrum of the
normal region is the solution of equation
\begin{align} \label{Dg}
\Delta_g = \tD(\Delta_g,\phi).
\end{align}
As follows from \Eq{tD}, at $\gamma\gg 1$ and $\phi=0$,  $\Delta_{g} \approx
\Delta/(1+\gamma)$.

In strongly asymmetric junctions with essentially different resistances of the
barriers, $R_{max} \gg R_{min}$, the transparency parameter $\gamma \approx
2\gamma_{min} = 2(R_{min}/R_N)(\Delta/E_{\textrm{Th}})$ is determined by the
smallest barrier strength. In this case, $\kappa \to 1$ and $\eta(\phi) \to 1$,
therefore the minigap weakly depends on the phase difference and approaches its
value at $\phi = 0$, while in the symmetric case the minigap oscillates with
the phase as $\Delta_g(0)|\cos(\phi/2)|$. The physical explanation is as
follows. In the main approximation, the stronger barrier can be considered as
impenetrable wall, therefore the spectrum of the N region, calculated using the
image method, is similar to the one for an effective SINIS junction with I
referring to the more transparent barrier, and N having doubled length (which
is manifested by doubled $\gamma_{min}$ in the estimate of $\Delta_g$). Since
both S electrodes in such an effective junction originate from the single S
electrode, the effective phase difference is zero within this approximation.

Expression for the current follows from \Eq{CommEq} in energy representation,
\begin{align} \label{defI}
I  = \int^\infty_{-\infty}\frac{dE}{16eR} \; \Tr\; \cG \; [\cG_-,\tau_K](E).
\end{align}
Using \Eqs{gequilib}-(\ref{tD}) and $G^K = (g^R-g^A)\tanh(E/2T)$, we get
\begin{align}
&I =  \frac{i \sin\phi}{4eR_T\eta(\phi)}\int^\infty_{-\infty} dE\;v^R \tv^R
\tanh\frac{E}{2T} - (R\to A),\nonumber
\end{align}
or in the Matsubara representation,
\begin{align} \label{I6}
I &= \frac{2\pi T}{eR_T} \sum_{\omega_n>0}\frac{1}{\sqrt{\omega_n^2
+\Delta^2}} \frac{\Delta^2 \sin \phi} {\sqrt{\omega_n^2 q_n^2 +\Delta^2
\eta^2(\phi)}},
\\
q_n &=1+\tau \sqrt{\omega_n^2 +\Delta^2},\quad i\omega_n = i\pi
T(2n+1).\nonumber
\end{align}

Equation \eqref{I6} coincides with the result of a direct solution of the
Usadel equation\cite{Kupr1999} and gives a general description for the
Josephson current in the double-barrier junctions.

At zero temperature, \Eq{I6} reduces to
\begin{align} \label{I8}
&I = \frac{\Delta \sin \phi}{eR_T}\times\begin{cases} \displaystyle
K\Bigl(\sqrt{ 1-\kappa^2}\sin\frac{\phi}{2}\Bigr), & \gamma\ll 1,
\\
\displaystyle\frac{1}{\gamma} \ln\frac{2\gamma}{\eta(\phi)}, & \gamma\gg
1.\end{cases}
\end{align}
where $K$ is the elliptic integral. These results have also been derived by
another methods for the chaotic quantum dot in ergodic regime\cite{Brouwer} and
for a diffusive junctions with equal\cite{KL,3T} and asymmetric\cite{Kupr1999}
barriers.

\section{Voltage biased Josephson junction}
\vskip -3mm

When the voltage is applied across the junction, the proximity state in the
normal region becomes nonstationary because of different time dependencies of
the electrode Green's functions in \Eq{g}. The periodicity of these functions
allows us to expand all matrices written in the $(E,t)$-representation over the
temporal harmonics, $A(E,t)=\sum\nolimits_m A(E,m) e^{-imeVt}$.

In this representation, the time averaged (dc) current $I = \overline{I(t)}$
reads
\begin{align} \label{I5}
I =\int_{-\infty}^\infty \int_{-\infty}^\infty  \frac{d\lambda\; dE}{16 \pi
eR} \sum_m \Tr \check{K}(\lambda,E,m) \bigl[\check{G}_-(E,-m), \tau_K \bigr].
\end{align}
Due to the fact that the local-equilibrium Green's functions in the electrodes,
\Eq{g}, contain only three harmonics, $m=0$ and $\pm1$, the current consists of
only three respective terms. By the same reason, equation for the matrix $\cK$
in \Eq{FormalSol}, $(\cA+i\lambda)\cK(\lambda)=1$, takes the form of the
three-term recurrency,
\begin{align}
&\left[\cG_+(E_m,0) -i\sigma_z \tau E_m +i\lambda\right] \cK_m
+\cG_+(E_{m-1/2},1)\cK_{m-1} \nonumber
\\
&+ \cG_+(E_{m+1/2},-1)\cK_{m+1} =\delta_{m,0},\label{EqK}
\end{align}
where $K_m(E) = K(E+meV/2, m)$ and $E_k=E+keV$.

In order to make the analysis of \Eqs{I5} and (\ref{EqK}) more tractable, we
perform in this Section some manipulations with the matrices $\cK_m$ and
$\cG_\pm$, in order to reveal the symmetries and simplify the structure of the
recurrence equation.

We start by introducing specific notations for the real-valued components of
the BCS Green's functions (\ref{gRL}),
\begin{align}
N=\re u^R,\quad M=\re v^R, \quad \oN=\im u^R,\quad \oM=\im v^R,
\\
(N,M)(E) \propto \theta(E^2-\Delta^2), \quad (\oN,\oM)(E) \propto
\theta(\Delta^2-E^2), \nonumber
\end{align}
where $\theta(x)$ is the Heaviside step function, $N(E)$ is the BCS density of
states, and write the functions $\cG_\pm$ explicitly,
\begin{align}
&\cG_\pm(E,m) = \sigma_z \delta_{m,0} \sum\nolimits_{\sigma=\pm}
\hat{t}^{\pm}_\sigma G_{0}^+(E_\sigma) +i\sigma_y \hat{t}^{\pm}_m G^+_1(E).
\label{Gpmexplicit}
\end{align}
Here we use the following abbreviations
\begin{align}
&G_{0}^+(E_\sigma)=\frac{1}{2}(i\oN_\sigma+N_\sigma F_\sigma), \;\;
G^+_{1}(E)= \frac{1}{2}(i\oM +MF), \label{G+}
\\
&\hat{t}^\pm_\sigma \equiv r_1 \hat{p}_\sigma \pm r_2 \hat{p}_{-\sigma}, \quad
F=\tau_z +2f\tau_+, \quad \sigma= \pm, \label{t}
\end{align}
where $E_\pm = E\pm eV/2$, $A_\pm =  A(E_\pm)$, $\tau_+ = (1/2)(\tau_x +
i\tau_y)$, $\hat{p}_\sigma = (1 + \sigma\sigma_z)/2$ are projectors in the
Nambu space, and the tensor products of the Nambu matrices $\hat{t}$ and
$2\times 2$ Keldysh matrices $G^+_{0,1}$ are assumed in \Eq{Gpmexplicit}. For
brevity, here and in the following we will avoid any special notations for such
matrices in the Keldysh space, keeping `check' for the $4 \times 4$ matrices
and `hat' for the $2\times 2$ Nambu matrices.

Equation (\ref{EqK}) can be presented in a more compact form,
\begin{align}
&\left(\check{Q}_m -i \tau E_m +i\sigma_z\lambda\right) \cK_m +\sigma_x
\left( \check{q}_{m-1} \cK_{m-1} + \check{q}'_m\cK_{m+1} \right) \nonumber
\\
&= \sigma_z\delta_{m,0}, \label{EqK2}
\end{align}
after multiplying (\ref{EqK}) by $\sigma_z$ and introducing notations
\begin{align}
&\check{Q}_m(E) = \hat{t}^+_+   H_m + \hat{t}^+_{-}   H_{m-1}, \quad H_m =
G_0^+(E_{m+1/2}) ,\label{Q}
\\
&\check{q}_m = \hat{t}^+_{+}   G_m, \quad \check{q}'_{m} = \hat{t}^+_{-}
G_m,\quad G_m = G^+_1(E_{m+1/2}). \label{q}
\end{align}
According to the definition of $\hat{t}^+_\sigma$ in \Eq{t}, the prime sign in
\Eq{q}  means the change $\sigma_z \to -\sigma_z$, or $\hat{p}_+
\leftrightarrow \hat{p}_{-}$, or $r_1 \leftrightarrow r_2$.

Now we show that the $4\times 4$ matrix recurrence \Eq{EqK2} can be simplified
and written in terms of the $2\times2$ matrices. Let us assume the ansatz
\begin{align}\label{ansatz}
&{\cK}_m = \begin{cases}  \sigma_x{\cP}_m \sigma_x{\cP}_{m-1}\ldots
\sigma_x{\cP}_1 {\cK}_0, & m>0,
\\
\sigma_x \cP_m \sigma_x \cP_{m+1}\ldots \sigma_x \cP_{-1} {\cK}_0, & m<0,
\end{cases}
\end{align}
which gives the recurrences for $\cP_m$, and also the expression for $K_0$ on
the form,
\begin{align}\label{sm}
&\cP_m = -\begin{cases}\left(\cQ'_m -i\tau E - i\sigma_z \lambda +\cq'_m
\cP'_{m+1}\right)^{-1}\cq_{m-1}, & m>0,
\\
\left(\cQ'_m -i\tau E - i\sigma_z \lambda +\cq_{m-1} \cP'_{m-1}
\right)^{-1}\cq'_m, & m<0, \end{cases}
\\
&\cK_0 =\sigma_z\left(\cQ_0 -i\tau E +i\sigma_z \lambda +\cq_0 \cP_1
+\cq'_{-1} \cP_{-1} \right)^{-1}.\label{K0}
\end{align}
According to \Eqs{t}, \eqref{Q}, and \eqref{q}, all quantities in \Eqs{sm} and
\eqref{K0} are diagonal in the Nambu space, and therefore these $4\times 4$
matrix relations split into a pair of $2\times 2$ separate relations for the
diagonal triangle Keldysh blocks $P_m$ and $K_0$ of the full $4\times 4$
matrices $\cP_m$ and $\cK_0$, respectively. These blocks differ one from
another by change of the sign of $\lambda$ (since $\lambda$ enters only through
the product $\sigma_z\lambda$) and by replacing $r_1 \leftrightarrow r_2$, in
accord with the structure of $\hat{t}_\sigma$.

Consider, for example, the upper block in the recurrences \Eq{sm} for $m>0$.
Denote $P'_m = P_m$ for $m=2k$, then
\begin{align}
&P_{1} = -(r_2 H_1 +r_1 H_0 -i\tau E - i \lambda +r_2 G_1 P_{2} )^{-1}r_1
G_0,\nonumber
\\
&P_{2} = -(r_1 H_2 +r_2 H_1 -i\tau E + i \lambda +r_1 G_2 P_{3} )^{-1}r_2
G_1,\quad\ldots\nonumber
\end{align}
We see that the recurrence coefficients with even index $m$ have the prefactor
$r_1$, while the coefficients with odd $m$ are multiplied by $r_2$. Thus,
introducing the notations
\begin{subequations}
\begin{align}
\rho_m &=  \begin{cases} r_1, & m=2k, \\ r_2, & m=2k+1,
\end{cases}\quad g_m=\rho_m G_m, \label{rho}
\\
h_m &=\rho_{m}H_m + \rho_{m-1}H_{m-1}-i\tau E_m+i(-1)^m\lambda ,
\\
\Pi_m &= \begin{cases} g_{m-1} P_m , & m>0, \\ g_{m} P_m , & m<0,
\end{cases}
\end{align}
\end{subequations}
and using a similar procedure for $m<0$, we finally arrive at the equation for
$2\times 2$ matrices $\Pi_m$,
\begin{align}
\Pi_m &= -\begin{cases} g_{m-1}(h_m  +\Pi_{m+1} )^{-1}g_{m-1}, & m>0,
\\ g_{m}(h_m  +\Pi_{m-1} )^{-1}g_{m}, & m<0, \end{cases},\label{Pim}
\\
K_0 &=(h_0 +\Pi_1  +\Pi_{-1} )^{-1}.\label{K0Pi}
\end{align}

Similar equation is valid for the lower Nambu block of the full Keldysh
matrices with the change $\Lambda \to -\Lambda$, where we introduce the
notation $\Lambda$ for the set $(\lambda,r_1,r_2)$ and $-\Lambda$ for
$(-\lambda,r_2,r_1)$.

Thus, in these terms, the three matrices $\cK(E,m)$, $m=0,\pm 1$, only relevant
in the dc current in \Eq{I5}, take the form,
\begin{align}
&\cK(E,0) =\cK_0(E) = \hat{p}_+   K_0(E,\Lambda) - \hat{p}_{-}
K_0(E,-\Lambda),\label{K0E}
\\
&\cK(E,\pm 1) =\cK_{\pm 1}(E_\mp)= \sigma_x (\cP_{\pm 1}\cK_0)(E_\mp),
\label{K(Epm1)}
\\
&\cP_m(E) =\hat{p}_+   P_m(E,\Lambda) + \hat{p}_{-} P_m(E,-\Lambda).\label{cP}
\end{align}
%

\section{Current-Voltage characteristics}
\vskip -3mm

As noted in previous Section, the current spectral density in \Eq{I5} can be
written as the sum of three terms,
\begin{align}
&\Tr\sum_{m=0,\pm1} \cK(E,m)\left[\cG_-(E,-m),\tau_K\right] =j_0+j_1+j_{-1},
\label{Jdens}
%
%
\\
&j_0=\Tr \, \cK_0(E) \sum\nolimits_{\sigma=\pm} \hat{t}^-_\sigma
G_0^-(E_\sigma),\label{j0}
\\
&j_{\pm 1} = - \Tr \; \cK(E,\pm 1)\sigma_x \hat{t}^-_{\mp}G_1^-(E)\nonumber
\\
&= -\Tr \; \hat{t}^-_{\pm }G_1^-(E_\pm)(\cP_{\pm 1}\cK_0)(E),\label{jpm1}
\\
&G_0^-(E_\sigma) = \tau_z N_\sigma f_\sigma + i\tau_y N_\sigma, \quad G_1^-(E)
= i\tau_x \oM +Mf. \nonumber
\end{align}
In \Eq{jpm1} we used \Eqs{K(Epm1)} and \eqref{ansatz}, then shifted the energy
by $\pm eV/2$ which holds the result of integration over energy in \Eq{I5}
unchanged. A direct calculation of the partial current density in \Eq{j0}
yields
\begin{align}
j_0 &=  r_1 N_+(2K_0^z f_+ -K_0^+) -r_2 N_-(2K_0^z f_-
-K_0^+)\nonumber
\\
&+ (\Lambda\to -\Lambda). \label{Iz}
\end{align}
Here and in the following, the upper indices $z$ and $+$ denote $\tau_z$- and
$\tau_+$-components of the Keldysh matrices, respectively. We note that the
change of sign of $\lambda$ in the last term to this equation plays no role
because of integration over $\lambda$ in \Eq{I5}; moreover, due to the
symmetries of the spectral functions in \Eq{Iz} with respect to $E \to -E$ (see
Appendix A), this term simply doubles the contribution of the upper line into
the full current.

Analysis of the contributions \Eq{jpm1} of the first harmonics performed in the
Appendix A [see \Eqs{j1}--\eqref{j_1x}] shows that all terms with unity
components of the matrices $K_0$ and $\Pi$ cancel each other after integration
over $E$ and $\lambda$ in \Eq{I5}. As the result, we finally arrive at the
following simplified expression for the dc current,
\begin{align}\label{I}
I&= \int_{-\infty}^\infty\int_{-\infty}^\infty\frac{dE\,d\lambda}{16\pi eR}\,
(j_0+j_1+j_{-1}),
\\
j_1+j_{-1}&=  \theta (\Delta^2-E_-^2)(K^+\Pi^z_{-1}-K^z\Pi^+_{-1}) \nonumber
\\
&- \theta (\Delta^2-E_+^2)(K^+\Pi^z_{ 1}-K^z\Pi^+_{ 1}) +(\Lambda \to
-\Lambda),\label{j1+j-1}
\end{align}
with $j_0$  given by \Eq{Iz}. As mentioned in comments to \Eq{Iz},  the change
$\Lambda \to -\Lambda$, due to integration over $\lambda$, can be reduced to
the permutation $r_1 \leftrightarrow r_2$.

\subsection{Numerical results}
\vskip -3mm

Numerical computation of current-voltage characteristics (IVCs) was done using
\Eqs{I} and \eqref{j1+j-1} with the function $K_0$ defined in \eqref{K0Pi} and
the solution $\Pi_{\pm 1}$  of the recurrence (\ref{Pim}). In this paper we
focus on the case, opposite to the one studied earlier,\cite{Our2011} of a
large difference between the barrier transparencies, say, $\gamma_1 \gg
\gamma_2$, where $\gamma_{1,2} = (R_{1,2}/R_N)(\Delta/E_{\textrm{Th}})$,
$\gamma^{-1} = (1/2)(\gamma^{-1}_1 + \gamma^{-1}_2)$. In this case, the
strongest barrier plays the role of a tunnel probe for the junction spectrum
formed basically by the weakest barrier, as was explained in comments to
\Eq{Dg}. On this account, we keep the relation $\gamma_1 = 10\gamma_2$, or,
equivalently, $R_1 = 10 R_2$ while calculating the IVCs at different
$\gamma_{1,2}$.

\begin{figure}[tb]
\centerline{\epsfxsize=8.5cm\epsffile{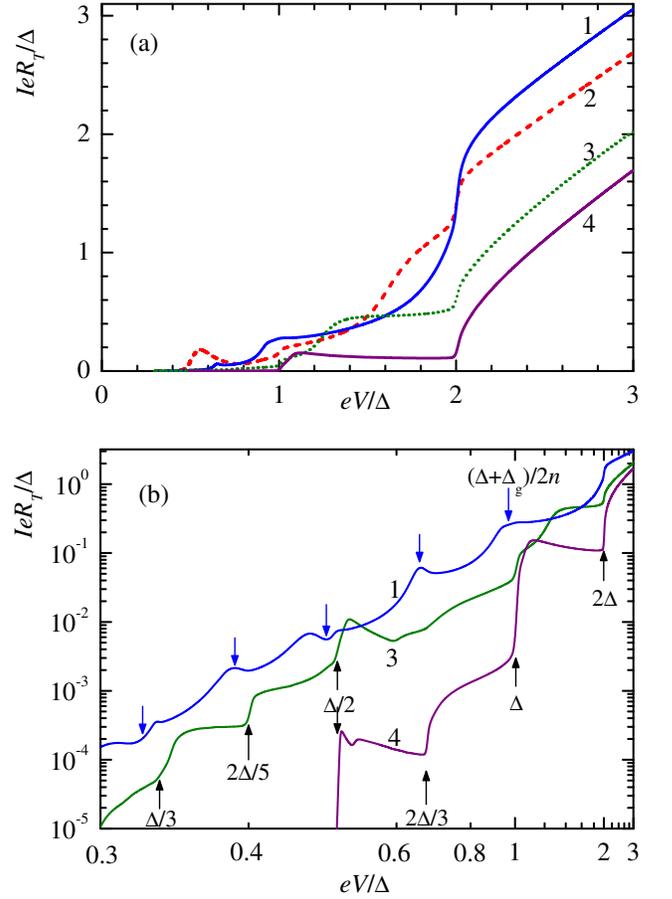}}
\vskip -5mm
\caption{(Color online) dc current vs applied voltage at $T=0$ for different
transparency parameters: $\gamma_2 = 0.1$, $\gamma_1 =1$, $\gamma = 0.18$,
$\Delta_g = 0.94\Delta$ (curve 1); $\gamma_2 = 0.5$, $\gamma_1 =5$, $\gamma =
0.91$, $\Delta_g = 0.57\Delta$ (curve 2); $\gamma_2 = 2$, $\gamma_1 =20$,
$\gamma = 3.64$, $\Delta_g = 0.22\Delta$ (curve 3); $\gamma_2 = 10$, $\gamma_1
=100$, $\gamma = 18.2$, $\Delta_g = 0.052\Delta$ (curve 4). (a) - current vs
voltage in linear scale, (b) - current in logarithmic scale vs voltage in
reciprocal scale. Downward arrows above the curve 1 indicate subharmonics of
$\Delta+\Delta_g$ for small values of $\gamma$.}
\vskip -6mm \label{sgs}
\end{figure}

In Fig.~\ref{sgs}, the results of numerical computation of the IVCs are shown
for several sets of $\gamma_{1,2}$. As one can see in Fig.~\ref{sgs}(a), the
excess current at large voltage is very small even at rather small $\gamma$,
and rapidly becomes negative, i.e., transforms to the deficit current, as long
as $\gamma$ increases. This is due to the strong asymmetry of the junction
assumed in our calculations, which confines the distribution of transparency
coefficients within the small enough interval, $0 < D < D_{\text{max}} \approx
0.4$. Such a  suppression of the excess current is similar to the case of a
junction with a single strong barrier.

For transparent barriers, $\gamma_2 = 0.1$, the IVC is close, as expected, to
the result of averaging of the current through a single-mode point contact over
the transparency distribution in a normal double-barrier structure, see
\Eqs{Nazarov} and \eqref{rhoD}. In this case, the steps in the IVC scale as
$(D_{\text{max}}/2)^{-2\Delta/eV}$; similar scaling has been found for the
tunnel junction with fixed transparency $D$ within the framework of
multiparticle tunneling theory \cite{MPT,MPT1} and MAR
theory.\cite{Bratus95,Averin95,Cuevas1996} The subharmonic features [shown by
downward arrows above the curve 1 in Fig.~\ref{sgs}(b)] are well fitted with
the ``combination'' subharmonics of the quantity $\Delta+\Delta_g$, although
they are quite close to the standard subharmonics of the bulk energy gap
$2\Delta$. The latter is explained by the fact that for transparent barriers,
the minigap $\Delta_g$ approaches $\Delta$.

\begin{figure}[tb]
\centerline{\epsfxsize=8.5cm\epsffile{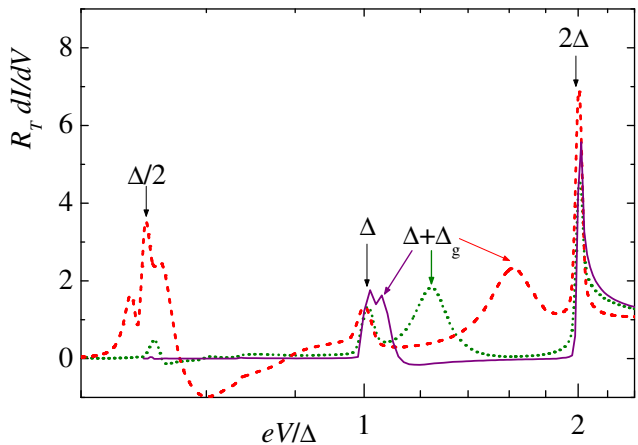}}
\vskip -3mm
\caption{(Color online) Differential conductance vs voltage at $T=0$: $\gamma_2
= 0.5$, $\gamma_1 =5$, $\Delta_g = 0.57\Delta$ (dashed curve); $\gamma_2 = 2$,
$\gamma_1 =20$, $\Delta_g = 0.22\Delta$ (dotted curve); $\gamma_2 = 10$,
$\gamma_1 =100$, $\Delta_g = 0.052\Delta$ (solid curve). The subharmonics of
the energy gap $2\Delta$ and the peaks at $eV = \Delta+\Delta_g$ are shown by
arrows.}
\vskip -6mm \label{deriv}
\end{figure}

With increasing barrier strengths, $\gamma_2 = 2-10$, the junction enters the
regime of strong dephasing, $\gamma \gg 1$. In this case, the role of the
effective tunneling parameter is played by $\gamma^{-1}$, as it was noted in
\cite{Our2011}, and, correspondingly, the IVC steps scale as
$\gamma^{-2\Delta/eV}$. This conclusion is confirmed by asymptotic analysis of
multiparticle currents presented in next subsection. The gap subharmonics
correspond to the current onsets [see Fig.~\ref{sgs}(b)], i.e., to maxima of
the differential conductance $dI/dV$. Such maxima are shown in
Fig.~\ref{deriv}, together with clearly pronounced peaks at
$eV=\Delta+\Delta_g$. The latter peaks are explained by enhanced transmissivity
of MAR chains containing links between the edges of the minigap and the bulk
gap where the density of state is enhanced. This effect is analogous to the one
in single channel resonant junctions,\cite{Johansson1999,Ingerman2001} where
additional peaks appear on IVC at voltages related to positions of geometric or
Andreev resonances in equilibrium.

Interestingly, similar splitting of the conductance peak near $eV =\Delta$ has
been found in Ref.~\onlinecite{Sam2002} for S-chaotic dot-S junction with the
minigap of the order of small Thouless energy $E_{\textrm{Th}} < \Delta$, which
corresponds to the long junction regime. The conductance peak at
$\Delta+\Delta_g$ has also been noted for an SNS junction with transparent
interfaces,\cite{Cuevas2006} and for a point contact (constriction) between
massive SN sandwiches.\cite{AZ} The physics in the latter case is similar to
the situation in asymmetric double-barrier junction considered in this paper:
the minigap is basically formed by the proximity effect in well-coupled S and N
regions, while the weak link, i.e., the constriction (in our case -- the
strongest barrier) plays a role of a probe, which weakly affects the spectrum
but detects its features in the IVC. Thus, the appearance of this specific
feature can be considered as a rather general phenomenon, which has also been
observed in experiments.\cite{Kutch1997,Kutch1999,Hoss2000} We note that the
strong asymmetry of the junction provides the most favorable conditions for
this effect: as noted above, in this case the minigap $\Delta_g(\phi)$
insignificantly depends on the superconducting phase difference $\phi(t)$ and
therefore holds nearly constant value $\Delta_g(0)$.

At very large $\gamma$ [curve 4 in Fig.~\ref{sgs}(a)], the minigap is small,
$\Delta_g \approx 0.05$, and therefore the splitting of the SGS at
$\Delta+\Delta_g$ remains visible only in the differential conductance while
the IVC features almost exactly correspond to the subharmonics of the
superconducting gap. In this case, the presence of the minigap manifests itself
in the IVC as anomalous enhancement of the magnitude of the dc current just
above the even gap subharmonics. This effect is due to the enhanced density of
states in the vicinity of the minigap which increases the transmissivity of the
MAR trajectories having even number of steps and therefore simultaneously
touching the superconducting gap edges and the small minigap region in the
middle of the bulk gap. This resonance effect becomes more pronounced for
higher subharmonics and leads to the appearance of the IVC portions with
negative differential conductance clearly visible in curves 3 and 4 in
Fig.~\ref{sgs}(b).

\subsection{Some analytical results}
\vskip -3mm

As it was mentioned, in junctions with a small dephasing parameter $\gamma\ll
1$, the problem reduces to the point contact limit,\cite{AZ} and eventually to
the single channel problem, which has been extensively
studied.\cite{Bratus95,Averin95,Cuevas1996} Here we present some analytical
results for the opposite limit of large dephasing $\gamma \gg 1$. In this case,
it is possible to express analytically the full dc current as a sum of
contributions of $n$-particle tunneling processes, similar to the single
channel theory.\cite{Bratus95} Solutions of the recurrence (\ref{Pim}) for the
quantities $\Pi_m$, which determine all functions necessary for the calculation
of the dc current in \Eq{I}, can be presented as perturbative expansion series
over the powers of $\gamma^{-1}$. Physically, these expansions reflect the
nature of the net current as a sum of $n$-par\-ticle tunnel currents; each of
them exists at $eV > 2\Delta/n$ and scales as $\gamma^{1-n}$ with respect to
the single-particle current. The latter fact allows us to consider the
$n$-particle current $I^{(n)}$ only within its actual voltage region $2\Delta/n
< eV < 2\Delta/(n-1)$; at larger voltages, the $(n-1)$-particle current
dominates. This simplifies further calculations and enables us to present the
net current in the form
\begin{align}\label{Iser}
&I  = \sum_{n=1}^\infty \chi_n(V)I^{(n)},
\\
& \chi_n(V)= \left\{ \begin{array}{ll} 1, & 2\Delta/n < eV < 2\Delta/(n-1);
\\ 0 & \textrm{otherwise} . \end{array}\right .\nonumber
\end{align}
Estimation shows that the $m$th term in the perturbative expansion for $\Pi_m$
contributes to the $(m+1)$-particle current; thus, it suffices to consider
them only at $eV < 2\Delta/m$, which greatly simplifies the structure of the
series.

We refer the reader to the Appendix B for the details of the evaluation of the
partial currents, which are rather cumbersome due to the junction asymmetry.
According to \Eq{In,T}, the $n$-particle current consists of $n$ equal
contributions of MAR chains with $n$ steps. Each chain starts at the energy $E
< -\Delta$ and finishes at $E>\Delta$, thus transferring the quasiparticles to
the extended states above the energy gap. The intermediate points in this chain
correspond to the energies inside the gap at which the Andreev reflections take
place. Here we present only final results for the first three partial currents
and the excess current.

The single-particle current exists at $eV>2\Delta$ and can be rather
straightforwardly evaluated for arbitrary temperatures,
\begin{align}\label{I1,T}
I^{(1)}&=  \int^{-\Delta +eV/2}_{\Delta -eV/2}
\frac{dE}{2e}\frac{N_+N_-(f_+-f_-)}{R_1N_++R_2N_-}+ \int^{\infty}_{\Delta
+eV/2}
\frac{dE}{2e} N_+N_-\nonumber \\
&\times  (f_+-f_-)\Bigl( \frac{1}{R_1N_++R_2N_-}+\frac{1}{R_2N_++R_1N_-}\Bigr)
\end{align}
(we remind that the subscripts $\pm$ denote the energy shift by $\pm eV/2$).

The spectral density of the 2-particle current calculated at $\Delta <eV<
2\Delta$ has a resonant form, with a sharp peak at zero energy. If the applied
voltage is not very close to the threshold $\Delta/e$ of the two-par\-ticle
current, the corresponding integral over energy can be calculated in the
resonant approximation. For simplicity, we present only the result for $T \ll
\Delta$,
\begin{align}\label{I2}
& I^{(2)}= \frac{\pi\Delta N(eV)}{2eR_T \gamma_1 \gamma_2} \sum_{i=1,2}
\frac{\gamma_i}{\sqrt{1+r_i^2N^2(eV)}}.
\end{align}

\begin{figure}[tb]
\centerline{\epsfxsize=8.5cm\epsffile{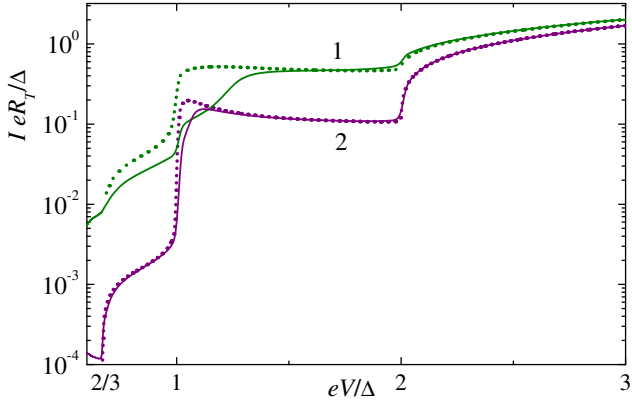}}
\vskip -3mm
\caption{(Color online) Comparison of the results of numerical calculation
(solid lines) and analytical approximation (dotted lines) for the contribution
of the 1-, 2-, and 3-particle currents to the net dc current: $\gamma_2 = 2$,
$\gamma_1 =20$ (curves 1); $\gamma_2 = 10$, $\gamma_1 =100$ (curves 2).}
\label{compar} \vskip -4mm
\end{figure}

The 3-particle current within the main approximation in $\gamma^{-1}$ at $T \ll
\Delta$ reads
\begin{align}\label{I3}
& I^{(3)}= \frac{3}{4e\gamma_1\gamma_2}\int^{-\Delta +3eV/2}_{\Delta -3eV/2}
\!\!\!\!\!\!\frac{N_{3/2}s_+s_-N_{-3/2}\; dE }{R_1s_-N_{-3/2}E^2_++R_2s_+N_{
3/2}E^2_- },
\end{align}
where $N_{\pm 3/2} = N(E\pm 3eV/2)$ and $s(E)=\oM^2(E)/4$. The numerator in
this equation clearly illustrates the structure of the relevant MAR chain: it
starts below the superconducting gap at the energy $E-3eV/2$, then the particle
experiences Andreev reflections inside the gap at the points $E \pm eV/2$, and
finishes above the gap, at the energy $E+3eV/2$. Figure \ref{compar}
demonstrates a rather good agreement between our purely numerical and
analytical results for the junction with opaque barriers, i.e., at large enough
barrier strength $\gamma$.

According to the definition, the excess current $I^{exc}$ is the
voltage-in\-de\-pen\-dent term in asymptotic expression for the dc current $I =
V/R_T + I^{exc}+O(\Delta/eV)$ at $eV \gg \Delta$. It is contributed by the
single-particle current and the two-particle (Andreev) current, and can be
evaluated for arbitrary $\gamma$, see Appendix C, where we restrict our
consideration to $T=0$. In the limit $\gamma\gg 1$, $I^{exc}$ appears to be
negative (deficit current), as one may expect for an opaque junction,
\begin{align}\label{excessgammainfty}
&I^{exc} =  \frac{-\Delta }{ eR_T} \Bigl[1+ \frac{1-a^2}{2a^{3/2}}\Bigl(
\frac{1}{2}\ln\frac{1+\sqrt{a}}{1-\sqrt{a}}-\arctan\sqrt{a}\Bigr)\Bigr],\\
&\gamma \gg 1, \quad a^2\equiv 1- r_1r_2, \nonumber \\
&I^{exc } = -\frac{\Delta }{eR_T}\begin{cases} {4}/{3}, & r_{1,2} =1 \;\;
\textrm{(symmetric junction)};
\\ 1, & r_1r_2\ll 1\;\;\textrm{(strong asymmetry)} .\end{cases}
\end{align}
For rather transparent interfaces, $\gamma \ll 1$, the excess current can be
expressed through the convolution of its value for a single ballistic
channel\cite{LTP97} with the transparency distribution \Eq{rhoD},
\begin{align}\label{excesstau=0}
I^{\textit{exc}} &=  \frac{\Delta }{ eR_T}\int_0^{D_{max}} \frac{D^2}
{\mathcal{R}}
\Bigl[1-\frac{D^2}{2\sqrt{\mathcal{R}(\mathcal{R}+1)}}\ln\frac{1+\sqrt{\mathcal{R}}}
{1-\sqrt{\mathcal{R}}}  \Bigr]
\\ \nonumber
&\times \rho(D)dD, \quad \mathcal{R} = 1-D.
\end{align}
%

\section{Conclusion}
\vskip -3mm

In conclusion, we have presented theoretical investigation of the
current-voltage characteristics in diffusive asymmetric SINIS Josephson
junctions with a short but finite length and different transparencies of SIN
interfaces. Our theory is relevant for current transport in Josephson devices
with multichannel semiconducting nanowires and multilayered planar metallic
junctions. We have shown that the coherent multiple Andreev reflection theory
can be efficiently developed and analyzed in detail, both numerically and
analytically, for the whole range of the interface transparencies and arbitrary
asymmetry.

We focused on the limit of resistive interfaces, when the dwell time of the
quasiparticles in the normal region becomes large, and the length dependence of
the transport characteristics becomes essential. Furthermore, we found that in
the case of significant asymmetry of the interface resistances, the subgap
current structures contain pronounced combination subharmonics of the bulk
energy gap and the proximity minigap, $\Delta+\Delta_g$, in addition to the
conventional subharmonics of the bulk energy gap $2\Delta$.  The effect of the
proximity minigap on the subgap current structures was found in a number of
numerical studies of various kinds of disordered SNS junctions, and also
observed in experiments. We argue, based on the detailed analytical study, that
this novel subgap structure is a robust feature and a general property of
diffusive SNS junctions.

In the limit of rather transparent interfaces, our theory recovers a known
formula for a short mesoscopic connector -- a convolution of the current
through a single-channel point contact with the transparency distribution for
an asymmetric double-barrier potential.

\begin{appendix}

\section{Symmetries}
\vskip -3mm

In this Section we discuss distinctive symmetries of the matrices $\Pi$ and
$K_0$, which allow us to simplify the expression for the dc current. For
simplicity, we first consider the case of equal barriers, $r_1=r_2=1$.
Beginning from the analysis of the symmetries of the matrix $h_m$, we will
assume in this Section all spectral and distribution functions to be dependent
on the energy $E+eV/2$, i.e., $N_m \equiv N[E+ eV(m+1/2)]$. This allows us to
write down the expansion of the matrix $h_m$ over the Pauli matrices in the
Keldysh space in the following form
\begin{align}\label{h}
h_m&=i[\oN_m + \oN_{m-1}-\tau E_m + (-1)^m \lambda]
\\
&+ \tau_z (N_m + N_{m-1}) + 2\tau_+ (N_m f_m + N_{m-1}f_{m-1}).\nonumber
\end{align}
Due to such indexing, the symmetry relations for the spectral and
distribution functions with respect to the change $E \to -E$ read as
\begin{align}
&N_m(-E) = N_{m'}(E),  & \oN_m(-E) = -\oN_{m'}(E),\nonumber
\\ \label{-E}
&M_m(-E) = -M_{m'}(E),  & \oM_m(-E) = \oM_{m'}(E),
\\
&f_m(-E) = -f_{m'}(E),  & m' = -m-1.\nonumber
\end{align}
By applying the transformation $(E,\lambda)\to(-E,-\lambda)$ to the function
$h_m$ in \Eq{h} and using \Eqs{-E}, we obtain the following relations for its
$1$-, $\tau_z$- and $\tau_+$-components denoted by corresponding upper
indices,
\begin{align}
&h_m^{1,+}(-E,-\lambda) = -h_{-m}^{1,+}(E,\lambda),\nonumber
\\
&h_m^z(-E,-\lambda) = h_{-m}^z(E,\lambda).\label{h-matrix}
\end{align}
In what follows, the Keldysh matrices with such symmetry properties will be
referred to as $h$-mat\-rices. It is easy to see that the inverse $h$-matrix is
a $h$-matrix too.

Now we will prove that the matrix
\begin{align} \label{pi}
&\pi_m(E,\lambda) = \begin{cases} g_{m-1}h_m^{-1}g_{m-1}, & m>0,
\\ g_{m}h_m^{-1}g_{m}, & m<0, \end{cases}
\end{align}
belongs to the class of $h$-matrices. By using the definitions of the functions
$g_m$ and $G_m$ in \Eqs{q} and \eqref{G+}, we get $g_m = (1/2)(i\oM_m + \tau_z
M_m) + f_m M_m \tau_+$. Denoting for brevity the $h$-matrix $h_m^{-1}$ as
$\oh_m$, we obtain at $m>0$
\begin{align} \label{m>0}
&\pi_m(E,\lambda)=g_{m-1}\oh_m g_{m-1} = -(1/4)\oM^2_{m-1} \oh_m \nonumber
\\
&+ (1/4)M_{m-1}^2[\oh_m^1 + \tau_z \oh_m^z + \tau_+(4f_{m-1}\oh_m^z -
\oh_m^+)].
\end{align}
Replacing $(E,\lambda)\to(-E,-\lambda)$ in \Eq{m>0}, using \Eqs{-E}, and
comparing the result with the expression for $\pi_{-m}(E,\lambda)$ with
negative indices,
\begin{align} \label{-m}
&\pi_{-m}(E,\lambda)=g_{-m}\oh_{-m} g_{-m} = -(1/4)\oM^2_{-m} \oh_{-m}
\nonumber
\\
&+ (1/4)M_{-m}^2[\oh_{-m}^1 + \tau_z \oh_{-m}^z + \tau_+(4f_{-m}\oh_{-m}^z -
\oh_{-m}^+)],
\end{align}
we see that the components of the matrix $\pi_m$ indeed satisfy
\Eqs{h-matrix}. Comparison of the definitions of the matrices $\Pi_m$ and
$\pi_m$, and the fact that the sum of $h$-matrices is the $h$-matrix too,
allows us to conclude that $\Pi_m$ is the $h$-matrix.

By using the expression \Eq{K0Pi} for the matrix $K_0$ through the matrices
$\Pi_{\pm 1}$ and the symmetry relations \Eqs{-E} and \eqref{h-matrix}, we see
that $K_0$ is the $h$-matrix with zero index, i.e.,
\begin{align} 
&K^{1,+}_0 (-E,-\lambda) = -K^{1,+}_0 (E,\lambda), \quad K^{z}_0 (-E,-\lambda)
= K^{z}_0 (E,\lambda).\nonumber
\end{align}

A generalization for the case of different barriers is rather obvious: since
the parity of indices of the spectral and distribution functions changes after
the transformation $E \to -E$ [see \Eqs{-E}], the symmetry relations
\Eqs{h-matrix} for the $h$-matrices must additionally involve the change $r_1
\leftrightarrow r_2$, in accordance with the definition \Eq{rho} of the
function $\rho_m$. In our notations, this is reduced to the substitution
$\lambda \to \Lambda$ in \Eqs{h-matrix}.

Now we consider the contribution of the first harmonics to the current density
\Eq{Jdens}. Using \Eqs{jpm1} and \eqref{K0E}--\eqref{cP}, we obtain
\begin{align}
j_1 &=-\Tr\; \hat{t}^-_+ G_1^-(E_+) \cP_1 \cK_0 \label{j1}
\\
&=2\Tr_\tau \left[ U_+\Pi_1 K_0 + (\Lambda\to -\Lambda)\right], \nonumber
\\
j_{-1} &=-2\Tr_\tau \left[ U_-\Pi_{-1} K_0 + (\Lambda\to -\Lambda)\right],
\label{j-1}
\\
U_\pm &= \frac{G_1^-(E_\pm) G_1^+(E_\pm)}{[G_1^+(E_\pm)]^2} \rightarrow
\frac{\tau_x \oM_\pm^2 - \tau_z M_\pm^2 f_\pm }{M_\pm^2-\oM_\pm^2}. \label{U}
\end{align}
Here we used the fact that $(G_1^+)^2$ is proportional to unity matrix and
omitted the terms with the matrix $\tau_+$, the trace of the product of which
with any triangle Keldysh matrix is zero.

It can be proved that all terms in the current spectral densities $j_{\pm
1}$, which contain unity matrix components, give no contribution to the full
dc current. Indeed, let us first consider the contributions of terms,
proportional to $\oM^2$:
\begin{align}
j_1^{\oM} &= -2\theta(\Delta^2 -E_+^2)\Tr_\tau \tau_x[\Pi_1 K_0 + (\Lambda
\to -\Lambda)] \label{j1x}
\\
&=-2\theta(\Delta^2 -E_+^2)[(\Pi_1^1+\Pi_1^z)K_0^+ +\Pi_1^+ (K_0^1 - K_0^z)
\nonumber
\\
&+ (\Lambda \to -\Lambda)], \nonumber
\\
j_{-1}^{\oM} &= 2\theta(\Delta^2 -E_-^2)\Tr_\tau \tau_x[\Pi_{-1} K_0+
(\Lambda \to -\Lambda)] \label{j_1x}
\\
&=2\theta(\Delta^2 -E_-^2)[(\Pi_{-1}^1+ \Pi_{-1}^z) K_0^+ +\Pi_{-1}^+ (K_0^1
-K_0^z)\nonumber
\\
&+ (\Lambda \to -\Lambda)].\nonumber
\end{align}
By using the symmetries \Eqs{h-matrix} of the $h$-matrices $\Pi$ and $K_0$, the
term $\theta(\Delta^2 -E_-^2)(\Pi_{-1}^1 K_0^+ +\Pi_{-1}^+ K_0^1)(E,\Lambda)$
in $j_{-1}^{\oM}$ can be transformed to the expression $[\theta(\Delta^2
-E_+^2) (\Pi_{1}^1 K_0^+ +\Pi_{1}^+ K_0^1)](-E,-\Lambda)$, which cancels the
analogous term with the arguments $(E,-\Lambda)$ in $j_{1}^{\oM}$ after
replacement $E \to -E$ in the integral in \Eq{I3}. Similar conclusions concern
the term $\theta(\Delta^2 -E_+^2)(\Pi_{1}^1 K_0^+ +\Pi_{1}^+ K_0^1)$ in
$j_{1}^{\oM}$ and the terms, proportional to $M^2$, all of which contain unity
matrix components.

\section{Analysis of partial multiparticle currents}
\vskip -3mm

Here we briefly describe the asymptotic analysis of these partial
contributions, using the methods and results developed earlier,\cite{Our2011}
with necessary modification due to asymmetry of the problem. To this end it is
useful to express all the relevant quantities through the following functions
\begin{align}
&{\mathcal N}_m=\rho_{ m}N_{m+1/2}\equiv \rho_{ m}N(E_{m+1/2}) ,\;\;\;
{\mathcal M}_m=\rho_m M_{m+1/2}
,\label{def} \\
&{\mathcal{\oN}}_m=\rho_{ m}\oN_{m+1/2},\quad {\mathcal{\oM}}_m=\rho_{
m}\oM_{m+1/2},\quad  \widetilde{f}_{m }=f_{m+1/2 }.\nonumber
\end{align}

Let us now express the current spectral density in \Eq{I},  $j_0+j_1+j_{-1}
\equiv 2j$, through the quantities introduced in \Eq{def} and put there
$r_1=r_2=1$ ($\rho_m=1$), which leads to the expression for the current
spectral density in the symmetric junction. Then we note that the derivation of
the asymptotic expressions of the current in the case of symmetric
junction\cite{Our2011} is performed by only using the analytical properties of
the functions in \Eqs{def} at $r_1=r_2$ and the symmetries with respect to the
permutation $E,\lambda \to -E,-\lambda$. Analysis shows that the functions for
the asymmetric junction defined in \Eqs{def} have the same analytical
properties and symmetries if we assume the simultaneous permutation
$r_1\leftrightarrow r_2$. Therefore we conclude that to obtain the expression
for the current in an asymmetric junction, one has to replace the current
spectral density of a symmetric junction as follows,
\begin{align}\nonumber
2j\{r_1=r_2\} \to j\{r_1,r_2\} + (r_1\leftrightarrow r_2), \quad R_1=R_2 \to R.
\end{align}
This enables us, using the results of Ref.~\onlinecite{Our2011}, to write down
the final formula for multiparticle currents in an asymmetric junction,
\begin{align}\label{In,T}
I^{(n)} &=  n(r_1 r_2)^n
\int_{-\infty}^\infty\frac{d\lambda}{2\pi}\int^{-\Delta
+(n-1/2)eV/2 }_{\Delta -eV/2}\frac{dE}{2eR}\; N_{1/2}N_{1/2-n}\nonumber \\
&\times  \Bigl[ \frac{1}{Z_0} \prod_{k=1}^{n-1}\frac{\overline
M^2_{1/2-k}}{4Z_{-k}} +(r_1\leftrightarrow r_2)\Bigr],
\\
\label{dets} Z_0 &= -\det (h_0  + \Pi_1 + \Pi_{-1}),
\\
\label{dets1} Z_{m\gtrless 0} &=-\det(h_{m}+ \Pi_{m \pm 1})
\end{align}
(at $n=1$, the product in \Eq{In,T} is assumed to be unity).

Practical calculations using \Eq{In,T} require an appropriate choice of
approximation for the determinants in \Eqs{dets} and \eqref{dets1}. These
quantities can be expressed, using the recurrence \Eq{Pim} for $\Pi_m$, through
the chain fractions that should be truncated at the $n$th step for the
$n$-particle current,
\begin{align}
&Z_m = |z_m|^2,
\\
&z_{m>0}= \widetilde{h}_m -\frac{g_{m}^2}{\widetilde{h}_{m+1}-
\frac{g_{m+1}^2}{\widetilde{h}_{m+2}-\ldots}}, \nonumber
\\
&z_{m<0}= \widetilde{h}_m -\frac{g_{m-1}^2}{\widetilde{h}_{m-1}-
\frac{g_{m-2}^2}{\widetilde{h}_{m-2}-\ldots}}, \nonumber
\\
&z_{0}=\widetilde{h}_0 -\frac{g_{0}^2}{\widetilde{h}_{1}-
\frac{g_{1}^2}{\widetilde{h}_{ 2}-\ldots}} - \frac{g_{ -1}^2}{\widetilde{h}_{
-1}-\frac{g_{ -2}^2}{\widetilde{h}_{ -2}-\ldots}}, \quad \widetilde{h}_m\equiv
h_m^1+ h_m^z, \nonumber
\end{align}
%

\section{Evaluation of excess current}
\vskip -3mm

The method of calculation of $I^{exc}$ is quite similar to that used in
Ref.~\onlinecite{Our2011}. The basic idea of this method relies on the fact
that only the energies of the order of $\Delta$ contribute into $I^{exc}$,
therefore at $eV \to \infty$ all spectral functions $M$, $\oM$, and $\oN$ with
``shifted'' energy $E+keV$, $k \neq 0$, turn to zero, and the density of states
$N(E+k eV)$ can be put to its limiting value (unity). This effectively
truncates the recurrences \Eq{Pim} for $\Pi_m$ and enables us to write down
$I^{exc}$ as the integral over $E$ and $\lambda$ of the explicitly defined
function. We will omit more detailed description of this procedure, which is
rather cumbersome due to the junction asymmetry, and present only final
results.

At arbitrary barrier strength $\gamma$, the integration over $\lambda$ can be
performed analytically which leads to the following expression at $T=0$,
\begin{align}\label{excess3}
&I^{exc}=  \frac{1 }{ eR_T} \Bigl( \int_{0}^{\Delta} dE\;j^< +
\int_{\Delta}^{\infty} dE\; N\; j^> \Bigr),
\end{align}
where
\begin{align}
& j^<  =  r_1r_2\oM^2 \sum_{i=1,2} (t_i\sqrt{t_i-c_i})^{-1},
\quad t_i=\sqrt{c_i^2+b_i^2}, \nonumber \\
& c_i=2(\tau^2E^2-1-\tau\oN r_i E) + r_1r_2, \nonumber\\
&  b_i=   2\tau E (2 -r_i)-r_1r_2\oN , \nonumber
\\
& j^> = \sum_{i=1,2}(T_i-A_-)^{-1/2}
( {A_+}/{T_i}+1)-2,\quad T_i=\sqrt{A_-^2+B_i^2},\nonumber \\
&A_\pm=2(\tau^2E^2\pm 1)\pm r_1r_2(N-1),\quad  B_i=2\tau E [ 2+r_i(N-1)].
\nonumber
\end{align}
At large $\gamma \gg 1$, the second term in \Eq{excess3} dominates, and the
integration can be done analytically, leading to \Eq{excessgammainfty}.

In the regime of small dephasing, $\gamma \ll 1$, it is reasonable to first
perform the integration over energy in the initial expression for $I^{exc}$ and
then, introducing the transparency variable $D= r_1 r_2 / (\lambda^2+1)$, to
express the excess current through its value for a single ballistic
channel\cite{LTP97} averaged over the transparency distribution \Eq{rhoD}, in
accordance with \Eq{Nazarov}, which results in \Eq{excesstau=0}.

\end{appendix}
\vskip -3mm


\end{document}